\documentclass[aps,prl,twocolumn,showpacs,floatfix,superscriptaddress]{revtex4-1}
\usepackage{graphicx,color}
\usepackage{times}
\usepackage{amsmath,amssymb}
\usepackage{multirow}
\usepackage{natbib}
\usepackage[colorlinks,citecolor=red]{hyperref}

\newcommand\hatH{{\hat{H}}}
\newcommand\hata{{\hat{a}}}
\newcommand\hatb{{\hat{b}}}

\newcommand\hatV{{\hat{V}}}
\newcommand\tila{{\tilde{a}}}

\newcommand\calO{{\mathcal{O}}}
\newcommand\calN{{\mathcal{N}}}
\newcommand\half{\frac{1}{2}}

\newcommand\ket[1]{\mathinner{\lvert{\textstyle#1}\rangle}}
\newcommand\braket[1]{\mathinner{\langle{\textstyle#1}\rangle}}

\definecolor{gray}{gray}{0.5}

\begin{document}

\title{The Kibble-Zurek Mechanism in a Topological Phase Transition}

\author{Minchul Lee}
\affiliation{Department of Applied Physics, College of Applied Science, Kyung Hee University, Yongin 446-701, Korea}

\author{Seungju Han}
\affiliation{Department of Physics, Korea University, Seoul 136-701, Korea}

\author{Mahn-Soo Choi}
\email{choims@korea.ac.kr}
\affiliation{Department of Physics, Korea University, Seoul 136-701, Korea}

\begin{abstract}
  The Kibble-Zurek mechanism (KZM) is generalized to a class of multi-level
  systems and applied to study the quenching dynamics of one-dimensional (1D)
  topological superconductors (TS) with open ends.
  Unlike the periodic boundary condition, the open boundary condition, that is
  crucial for the zero-mode Majorana states localized at the boundaries,
  requires to consider many coupled levels.
  Our generalized KZM predictions agree well with the numerically exact results
  for the 1D TS.
\end{abstract}

\pacs{64.60.Ht, 11.15.Ha, 73.20.At, 73.43.Nq}

\maketitle

Conventional second-order phase transitions (PTs) are driven by the spontaneous symmetry breaking and typically described by local order parameters, which take continuous values. Various critical scalings are traced back to symmetry breaking.
To the contrary, topological PTs involves the change in internal topology rather than symmetry breaking. Necessarily, topological states are classified by topological quantum numbers, which are discrete.
For instance, topological insulators and superconductors are characterized by the number of gapless boundary (surface, edge or endpoint) states \cite{Hasan10a,Qi11a,Shen12a} separated from gapped bulk states.
These observations raise an intriguing question of how the topological order emerges or disappears temporally when system parameters are quenched across the critical point \cite{Perfetto13a,DeGottardi11a,Bermudez10a,Bermudez09a}.

This question comes up ever more curious when one recalls that the Kibble-Zurek mechanism (KZM), a theory of the formation of topological defects in second-order PTs, establishes quite accurate connections between the equilibrium critical scalings and the nonequilibrium dynamics of symmetry breaking.
The KZM was originally put forward to study the cosmological PT of the early Universe \cite{Kibble76a,Kibble80a} and later extended to study classical PTs in condensed matters \cite{Zurek85a,Zurek96a}. Recently, it was found to apply to the Landau-Zener transitions in two-level quantum systems \cite{Damski05a,Damski06a} and the dynamics of second-order quantum PTs \cite{Dziarmaga05a,Dziarmaga12a} as well.
In fact, these latter two classes of dynamics share a key characteristic, the ``critical slowing down'' which comes from the critical scaling of the correlation length for the former and the reduced level spacing for the latter.
Nevertheless, the agreement between the KZM prediction and the exact dynamics still remains ``somewhat surprising'' \cite{Zurek05a}.
It is then a demanding question whether topological PTs, which are not even driven by symmetry breaking (critical scaling), can be described in the spirit of KZM.
Interestingly, a recent study of the Creutz ladder and the $p$-wave superconductor wire pointed out that topology makes the density of defects deviate strongly from the \emph{two-level} KZM scaling \cite{Bermudez09a,Bermudez10a}.

In this work, we generalize the KZM to a class of multi-level systems and apply it to study the quenching dynamics of one-dimensional (1D) topological superconductors (TS).
We stress that the open boundary condition (OBC), which is crucial for the zero-mode Majorana states localized at the boundaries, requires us to consider many coupled levels \cite{Bermudez09a}. 
Under the periodic boundary condition (PBC), the system is essentially a two-level system involving two modes of opposite momenta \cite{Demkov95a,Dziarmaga05a}.
To extend the KZM to multi-level systems we formulate the dynamics using the
dynamical transition matrix and develop the so-called conserving and
non-conserving KZM. Both are equivalent to the KZM for two-level systems.  Our
generalized KZM predictions, taking into account the Majorana states formed at
its ends and its dynamical transition into multi-levels, agree well with the numerically exact results for the 1D TS.
Our new approach may provide an insight of the surprisingly good agreement between the KZM and the exact dynamics in two-level systems, and shed light on possible extensions of the method to even more general classes of systems.


\paragraph{Model.}

A 1D TS of length $L$ is described by
the tight-binding Hamiltonian of spinless fermions \cite{Kitaev01a}
\begin{equation}
\label{Paper::eq:1}
\hat{H}(t)
= \frac{w}{2}
\sum_{j=1}^{L-1}
\left[\hat{c}_j \hat{c}_{j+1} - \hat{c}_j^\dag \hat{c}_{j+1} + h.c.\right]
- \mu(t) \sum_{j=1}^{L}\hat{c}_j^\dag \hat{c}_j.
\end{equation}
Here, for simplicity, we take the Ising limit in which the $p$-wave
superconducting order parameter $\Delta$ is equal to the hopping amplitude
$w$.
In the quenching process, the chemical potential
\begin{math}
\mu(t)L = w(t/\tau_Q+L)
\end{math}\cite{endnote:4}
is ramped up from $0$ to $\infty$ through the transition point $\mu=w$ at
$t=0$.  The process drives the system from the topological ($|\mu|<w$)
to trivial ($|\mu|>w$) phase.

In the continuum limit, Eq.~(\ref{Paper::eq:1}) is reduced to the Dirac Hamiltonian
\begin{equation}
\label{Paper::eq:2}
\hatH(t)
= \half\int{dx}\,\hat\Psi^\dag(x) H(x,t) \hat\Psi(x) \,,\quad
\hat\Psi =
\begin{bmatrix}
\hat\psi \\
\hat\psi^\dag
\end{bmatrix}
\end{equation}
with
\begin{math}
H(x,t) = M(x,t)v_s^2\tau_z - i\hbar{v_s}\tau_x\partial_x,
\end{math}
where $\tau_x,\tau_y,\tau_z$ are the Pauli matrices in the particle-hole space,
$\hbar{v_s}=a\Delta$ with $a$ being the lattice constant, and $Mv_s^2=\mu-w$.
Hereafter we use the unit system such that $\hbar=v_s=a=1$.
The position-dependent ``mass'' $M(x,t)$ accounts for the spatially inhomogeneous regions of the TS. We are particularly interested in the case \cite{Fu08a,endnote:3}
\begin{equation}
M(x,t) =
\begin{cases}
\infty & (|x|>L/2) \\
M(t) & (|x|\leq L/2)
\end{cases}.
\end{equation}
When $M(t)<0$, there exist two zero-energy Majorana Fermions localized at
$x=\pm L/2$ \cite{Shen12a}. In the continuum limit, we consider the quenching
of the form $M(t)L=t/\tau_Q$ \cite{endnote:4}. For simplicity we mostly
discuss the dynamics in terms of the continuum model; qualitative features are
the same.

\begin{figure}
\centering
\includegraphics[width=4cm]{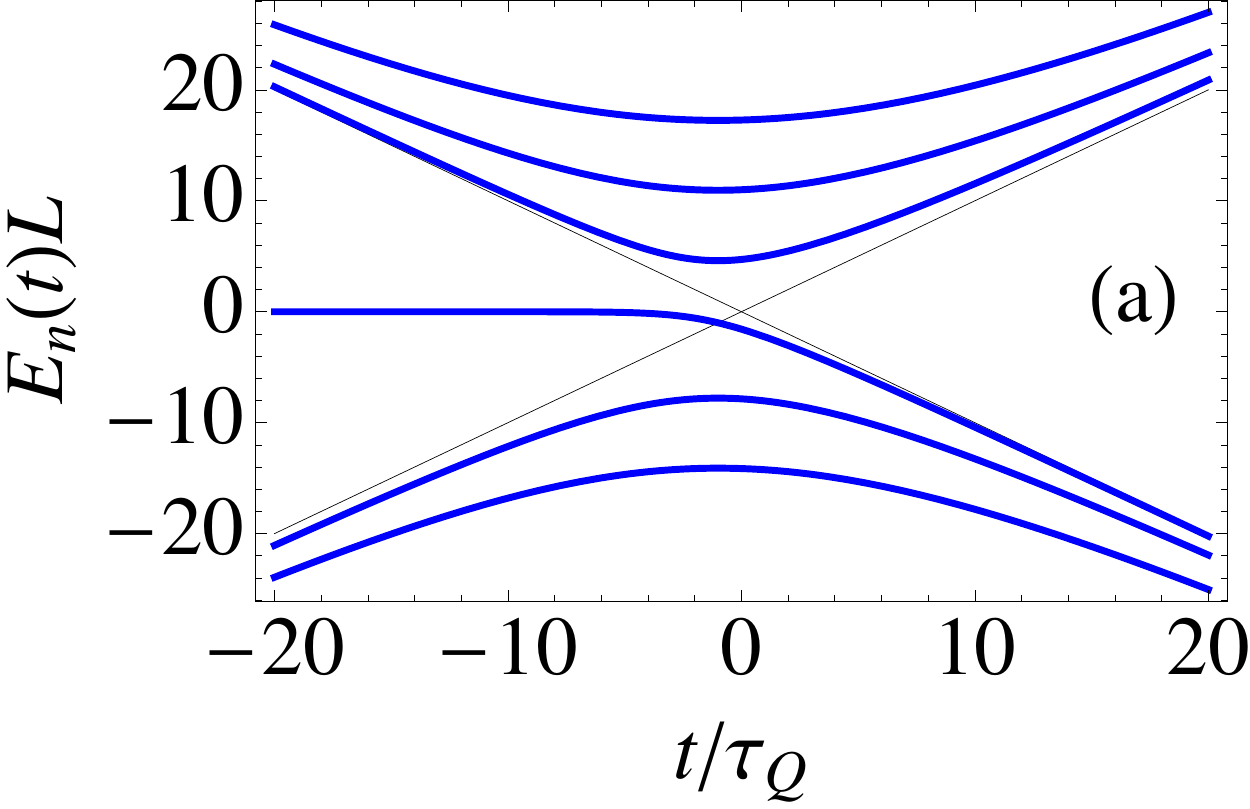}
\includegraphics[width=4cm]{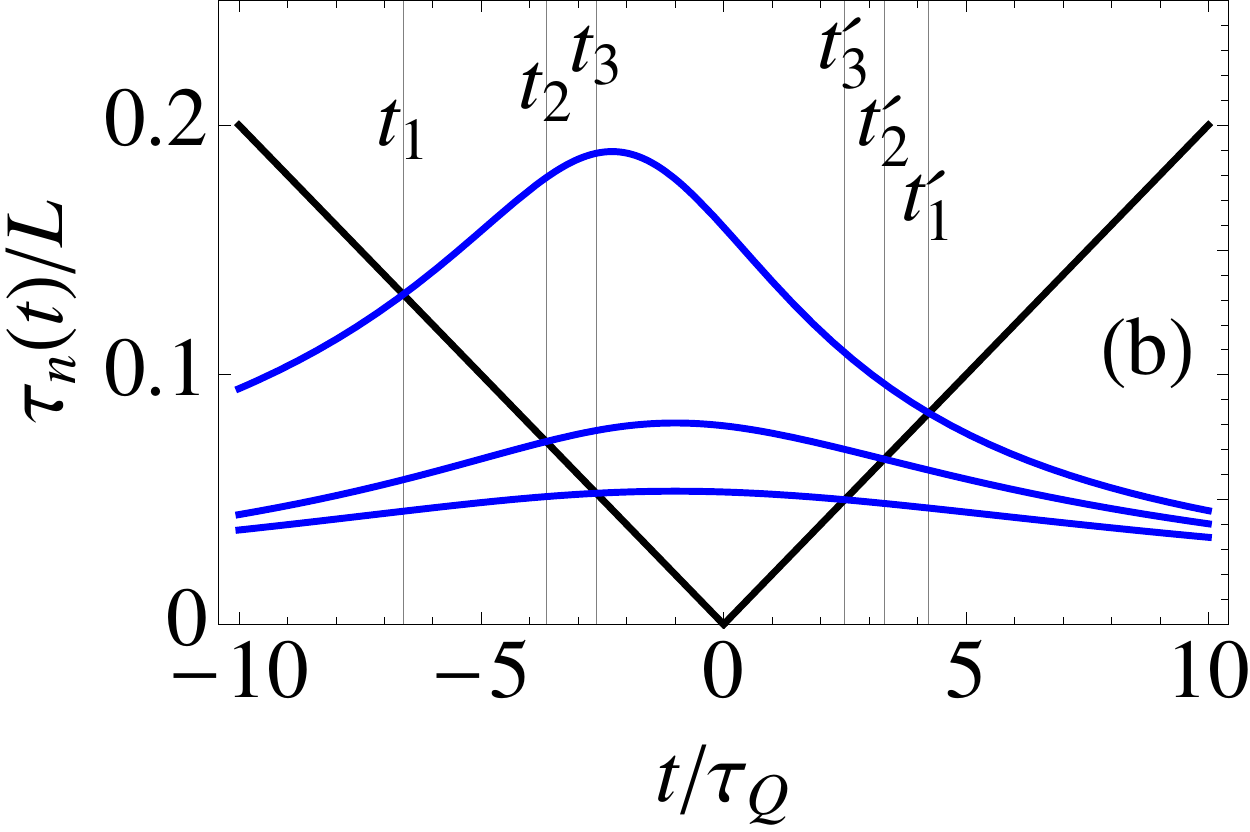}
\caption{(a) The quasi-particle energy levels (collecting only even-parity
  modes) and (b) relaxation time scales in a 1D TS of length $L$ and ``mass''
  $M(t)L=t/\tau_Q$, where $\tau_Q$ is the quenching time. The
  AI crossover points $t_n$ and $t_n'$ ($n=1,2,\cdots$) are
  indicated by thin vertical lines in panel (b).}
\label{Paper::fig:1}
\end{figure}

We start with the single-particle Dirac equation
\begin{equation}
\label{Paper::eq:3}
H(x,t)\Phi_n(x,t) = E_n(t)\Phi_n(x,t).
\end{equation}
It has two important symmetries, the space inversion and the particle-hole
symmetry.  The inversion symmetry allows us to choose solution to be parity
eigenstate and subject to the boundary conditions
\begin{equation}
\Phi_n(L/2,t) = \pm\tau_z\Phi_n(-L/2,t) =
\begin{bmatrix}
1 \\ i
\end{bmatrix},
\end{equation}
where the sign $\pm$ corresponds to the even/odd parity under the space
inversion.
Because of the particle-hole symmetry, if $\Phi_n(x,t)$ is a solution of the
Dirac equation with energy $E_n(t)$, then its charge conjugation partner
$\tau_x\Phi_n^*(x,t)$ is also a solution but with energy $-E_n(t)$.
Further, if $\Phi_n(x,t)$ has a definite (even or odd) parity, then
$\tau_x\Phi_n^*(x,t)$ has the opposite parity.  Hence it suffices to count
only, say, even-parity solutions.
Hereafter we reserve the notation $\Phi_n(x,t)$ for the even-parity modes,
\begin{equation}
\Phi_n(x,t) =
\begin{bmatrix}
\sin(k_nL/2)\cos(k_nx) \\
i\cos(k_nL/2)\sin(k_nx)
\end{bmatrix}
\end{equation}
with $k_n(t)$ satisfying $\tan(k_nL)=-k_n/M$.  Odd-party modes are referred to
by $\tau_x\Phi_n^*(x,t)$.  The mode $\Phi_n(x,t)$ has energy $E_n(t) =
(-1)^{n-1}\sqrt{M^2(t)+k_n^2(t)}$ with $n=0,1,2\cdots$ in the increasing order of $|E_n|$, whose time dependence is illustrated in
Fig.~\ref{Paper::fig:1}(a).  Of particular importance is the \emph{zero-mode},
$\Phi_0(x,t)$, whose energy $E_0(t)\approx M(t)/\cosh(M(t))$ is exponentially
small for $M(t)L<-1$ and physically responsible for the Majorana modes
localized at the interfaces ($k_0$ is purely imaginary).

We denote the quasi-particle operator for the mode $\Phi_n(x,t)$ and $\tau_x\Phi_n^*(x,t)$ by $\hata_n(t)$ and $\hatb_n(t)$, respectively. Obviously, $a_n^\dag(t) = b_n(t)$.
In terms of these the many-body Dirac Hamiltonian~(\ref{Paper::eq:2}) reads
\begin{math}
\hatH(t)
= \sum_{n=0}^\infty E_n(t)\left[\hata_n^\dag(t)\hata_n(t)-1/2\right].
\end{math}
The actual dynamics is governed by the Heisenberg operators $\tila_n(t)$,
related to the instantaneous eigenoperators $\hata_n(t)$ by
\begin{math}
\tila_n(t) = \hatV^\dag(t)\hata_n(t)\hatV(t)
\end{math}
where $\hatV(t)$ is the many-body time-evolution operator
\begin{math}
\hatV(t) = T\exp\left[-i\int_{-\infty}^t{ds}\,\hatH(s)\right].
\end{math}
They satisfy the Heisenberg equation of motion
\begin{equation}
\label{Paper::eq:4}
i\frac{d}{dt}\tila_m(t) = \sum_n K_{mn}(t)\tila_n(t)
\end{equation}
with
\begin{math}
K_{mn}(t) \equiv \delta_{mn}E_n(t) - \Omega_{mn}(t)
\end{math}
and
\begin{math}
\Omega_{mn}(t) \equiv i\braket{\Phi_m(t)|\dot\Phi_n(t)}
\end{math}
or, equivalently,
\begin{equation}
\label{Paper::eq:5}
\tila_m(t)=\sum_nU_{mn}(t,t_0)\tila_n(t_0)
\end{equation}
with $U(t,t')=T\exp\left[-i\int_{t'}^t{ds}\,K(s)\right]$.
The many-body dynamics in Eq.~(\ref{Paper::eq:4}) is intimately related to the single-particle dynamics: When a wave function $\ket{\Psi(t)}$ is expanded into
\begin{math}
\ket{\Psi(t)} = \sum_n\beta_n(t)\ket{\Phi_n(t)}
\end{math}
the amplitudes $\beta_n(t)$ satisfy the effective Schr\"odinger equation
\begin{equation}
\label{Paper::eq:6}
i\frac{d}{dt}\beta_m(t)
= \sum_nK_{mn}(t)\beta_n(t)
\end{equation}
or, equivalently,
\begin{equation}
\beta_m(t)=\sum_nU_{mn}(t,t_0)\beta_n(t_0).
\end{equation}
The effective Hamiltonian $K(t)$ in~(\ref{Paper::eq:4}) and (\ref{Paper::eq:6}) includes off-diagonal elements $\Omega_{mn}(t)$
with the common phase fixing choice
\begin{math}
\Omega_{nn}(t) = 0.
\end{math}
Mathematically, the matrix $\Omega(t)$ gives the \emph{dynamical connection} between the instantaneous eigenstates at different times,
\begin{math}
\braket{\Phi_m(t)|\Phi_n(t')} = W_{mn}(t,t')
\equiv T\exp\left[i\int_{t'}^t{ds}\,\Omega(s)\right].
\end{math}
Physically, $\Omega_{mn}(t)$ is responsible for the \emph{dynamical transitions} between different instantaneous energy levels $E_m(t)$ and $E_n(t)$.

The energy levels $E_n(t)$ and the dynamical transitions $\Omega_{mn}(t)$
between them in 1D TS [see Fig.~\ref{Paper::fig:1}(a)] have peculiar
properties: The level spacings satisfy
\begin{equation}
\label{Paper::eq:7}
|E_{n-1}(t)-E_n(t)| < |E_n(t)-E_{n+1}(t)|
\end{equation}
and the direct transition is allowed only for nearest-neighbor pairs of levels
\begin{equation}
\label{Paper::eq:8}
\Omega_{mn}(t)\approx 0 \text{ unless } m=n\pm1 \,.
\end{equation}
These two properties are pivotal in our generalization of the KZM below.

In passing, the property~(\ref{Paper::eq:8}) casts a sharp contrast between the
OBC and PBC \cite{Dziarmaga05a}. Under the PBC, momentum is conserved and
transitions occur only between modes with opposite momenta $k$ and $-k$:
$\Omega_{kk'}=0$ unless $k+k'=0$. Therefore, the dynamical model is essentially
a two-level system \cite{Demkov95a,Dziarmaga05a} and the KZM for two-level
systems is enough. Of course, in the thermodynamic limit, the boundary
condition does not make difference in bulk states. However, the Majorana states
at the boundaries do not have a counterpart under the PBC and cause the
inherently multi-level dynamics.

\paragraph{Generalized Kibble-Zurek Mechanism.}

Let us first consider a single-particle dynamics by taking into account $N+1$
levels ($N=\infty$ for the continuum model).  We
suppose that the system was initially in the $n=0$ instantaneous eigenstate,
say,
\begin{math}
\ket{\Psi(t=-\infty)}=\ket{\Phi_0(-\infty)}
\end{math}
and examine the final state $\ket{\Psi(\infty)}$ in the far future.
Within the spirit of the KZM \cite{Zurek05a,Damski05a,Damski06a}, we determine \emph{adiabatic-impulse (AI) crossover points} $t_n$ and $t_n'$ by comparing the \emph{relaxation time scale}, $\tau_n(t)=1/|E_n(t)-E_{n-1}(t)|$, and the time scale for the relative coupling to develop, $M(t)/\dot{M}(t)=t$:
\begin{equation}
\label{Paper::eq:15}
\tau_n(t_n) = -\alpha t_n \,,\;\;
\tau_n(t_n') = +\alpha t_n' \;\; (1\leq n\leq N),
\end{equation}
where $\alpha=\calO(1)$ is a fitting parameter \cite{endnote:1}.  Due to the
level-spacing structure in Eq.~(\ref{Paper::eq:7}), the crossover points are
arranged in the order
\begin{math}
t_1 < \cdots < t_N < t_N' < \cdots < t_1'
\end{math}
[see Fig.~\ref{Paper::fig:1}(b)]. Here note that the crossover points are \emph{not} symmetric about the critical point ($t_n\neq -t_n'$) \cite{Damski06a}.
The asymmetry is due to the Majorana modes, which exist only for $M(t)L<-1$.

The initial evolution from $t=-\infty$ to $t_1$ is completely adiabatic and thus
\begin{math}
\ket{\Psi(t_1)} = \ket{\Phi_0(t_1)}.
\end{math}
From this moment to $t_2$, the two levels $E_0(t)$ and $E_1(t)$ become
impulsive but the rest, far away from the two, still remain unpopulated. In the
two-level case, the AI approximation assumes that the state remains completely
intact: $\ket{\Psi(t_2)}=\ket{\Psi(t_1)}$. A vital difference in the
multi-level case is that it violates the probability conservation
because even the relatively adiabatic states $\ket{\Phi_n(t_2)}$ ($n\geq 2$) have finite overlaps with $\ket{\Phi_0(t_1)}$ and
$\ket{\Phi_1(t_1)}$.

Therefore, we instead adopt to ``prune'' the effective Hamiltonian $K_{mn}(t)$ as following: Suppose that the first $(r+1)$ levels $E_0,E_1,\cdots,E_r$ are impulse. Then we ignore the energy differences among impulse levels, $E_m\approx 0$ ($0\leq m\leq r$), and keep only the the dynamic transitions between impulse levels, $\Omega_{mn}(t)\approx0$ either for $0\leq m\leq r<n$ or for $r<m,n$. This leads to the \emph{pruned} effective Hamiltonian
\begin{equation}
\label{Paper::eq:9}
K^{(r)}_{mn}(t) =
\begin{cases}
-\Omega_{mn}(t) & (m,n\leq r) \\
\delta_{mn}E_n(t) & (\text{otherwise})
\end{cases}
\end{equation}
and the corresponding pruned evolution matrix
\begin{math}
U^{(r)}(t,t') \equiv
T\exp\left[-i\int_{t'}^t{ds}\,K^{(r)}(s)\right].
\end{math}
The pruning amounts to evolving the impulse levels solely by the dynamic transition matrix $\Omega_{ij}$ while keeping the adiabatic levels intact. Being unitary, $U_{mn}^{(r)}$ preserves the probability. Within the AI approximation, the evolution is thus expected to be governed by
\begin{multline}
U^{(r,s)} \equiv
U^{(r)}(t_r',t_{r+1}')\cdots U^{(N-1)}(t_{N-1}',t_N')\times{} \\{}
U^{(N)}(t_N',t_N) U^{(N-1)}(t_N,t_{N-1})\cdots
U^{(s)}(t_{s+1},t_s).
\end{multline}

Indeed, getting back to the example, the evolution from $t_1$ to $t_2$ is described by
\begin{math}
\beta_m(t_2) \approx \sum_n U^{(1)}_{mn}(t_2,t_1)\beta_n(t_1).
\end{math}
Note that for the two-level case ($N=1$) this is equivalent to the original AI approximation \cite{Damski05a,Damski06a}. The same procedure is repeated until $t_N$ to get (recall $\beta_{n}(-\infty)=\delta_{n0}$)
\begin{equation}
\beta_m(t_N) = \Big[U^{(N-1)}(t_N,t_{N-1})\cdots U^{(1)}(t_2,t_1)\Big]_{m0}.
\end{equation}
After the moment $t=t_N'$, the level $E_N(t)$ becomes \emph{relatively}
adiabatic again and its occupation probability does not change from
$|\beta_N(t_N')|^2$. The rest evolve impulsively until $t=t_{N-1}'$, when the
level $E_{N-1}(t)$ becomes relatively adiabatic. Repeating this approximation
until $t=t_1'$, after which the whole evolution becomes adiabatic, one finally
obtains the AI approximation for the amplitudes
\begin{math}
\beta_m(\infty) \approx U^{(m,n)}_{m0} \,.
\end{math}
Similarly, starting from a general initial state $\ket{\Phi_n(-\infty)}$ with $n>0$ one gets the occupation probabilities
\begin{align}
P_{m|n}(\infty)
&\approx \left|U^{(m,n)}_{mn}\right|^2,&
P_{m|0}(\infty)
&\approx \left|U^{(m,1)}_{m0}\right|^2, \nonumber \\
P_{0|0}(\infty)
&\approx \left|U^{(1,1)}_{00}\right|^2,&
P_{0|n}(\infty)
&\approx \left|U^{(1,n)}_{0n}\right|^2,
\label{Paper::eq:10}
\end{align}
where $m,n>0$.
Equation~(\ref{Paper::eq:10}) is called the \emph{conserving KZM} for the
multi-level system as it conserves the probability, $\sum_mP_{m|n}(\infty)=1$. It generalizes the KZM for two-level systems, and the
calculation involves simple procedures requiring only instantaneous
eigenvectors.

Although the expression~(\ref{Paper::eq:10}) requires only instantaneous eigenvectors at discrete times, one still needs to calculate the time-ordered exponential function of the matrix $\Omega(t)$. As we will see now, in many cases it can be avoided.
For a large system, the AI crossover points are closely packed
and each factor in~(\ref{Paper::eq:10}) can be approximated by
\begin{equation}
\label{Paper::eq:11}
U^{(r)}_{ij}(t+\eta,t) \approx
\begin{cases}
\delta_{ij} + i\eta\Omega_{ij}(t) & (i,j\leq r) \\
\delta_{ij}[1-i\eta E_j(t)] & (\text{otherwise})
\end{cases}
\end{equation}
up to $\calO(\eta^2)$.  When Eq.~(\ref{Paper::eq:11}) is substituted into
Eq.~(\ref{Paper::eq:10}), due to Eq.~(\ref{Paper::eq:8}), only the subpart
$\delta_{ij}+i\eta\Omega_{ij}(t)$ ($i,j\leq r$) of each matrix
$U^{(r)}(t+\eta,t)$ contribute to the product; hence $U^{(r)}(t+\eta,t)$ in
Eq.~(\ref{Paper::eq:10}) can be replaced safely with $1+i\eta\Omega(t)\approx
W(t+\eta,t)$ up to $\calO(\eta^2)$. Then the probability reduces to
[recall that $W_{mn}(t',t)=\braket{\Phi_m(t')|\Phi_n(t)}$]
\begin{subequations}
\label{Paper::eq:12}
\begin{align}
P_{m|n}(\infty)
& \approx 
\left|\braket{\Phi_m(t_m')|\Phi_n(t_n)}\right|^2, \\
\label{Paper::eq:16}
P_{m|0}(\infty)
&\approx \left|\braket{\Phi_m(t_m')|\Phi_0(t_1)}\right|^2 \,,\\
P_{0|n}(\infty)
&\approx \left|\braket{\Phi_0(t_1')|\Phi_n(t_n)}\right|^2 \,,\\
P_{0|0}(\infty)
&\approx \left|\braket{\Phi_0(t_1')|\Phi_0(t_1)}\right|^2 \,,
\end{align}
\end{subequations}
where $m,n>0$.
This approximation, which we call the \emph{non-conserving KZM}
for the multi-level system, drastically simplifies the calculation of
$P_{m|n}$ which demands only the overlap integrals of instantaneous
eigenvectors at different times.
The caveat is that it violates the probability conservation (hence the name ``non-conserving''), $\sum_mP_{m|n}(t)<1$, as it involves eigenstates $\ket{\Phi_m(t_m')}$ at different times for different levels. The amount of violation, $\varepsilon=1-\sum_mP_{m|n}(\infty)$, gives a convenient estimate of the error.
The result~(\ref{Paper::eq:12}) implies that given the initial state $\ket{\Phi_n(-\infty)}$ the system essentially remains impulse from $t_n$ to $t_m'$. Indeed, the non-conserving KZM essentially assumes that the part associated with the relatively impulse levels remains completely intact (see the discussion above Eq.~(\ref{Paper::eq:9})).
However, the derivation of the non-conserving KZM via the conserving KZM using the pruned evolution matrix $U^{(r)}_{mn}$ paves a way to further generalizations of the KZM for systems with more complicated level and coupling structure.
Moreover, in practice, the violation does not affect its accuracy much as demonstrated in below.

\begin{figure}
\centering
\includegraphics[width=8cm]{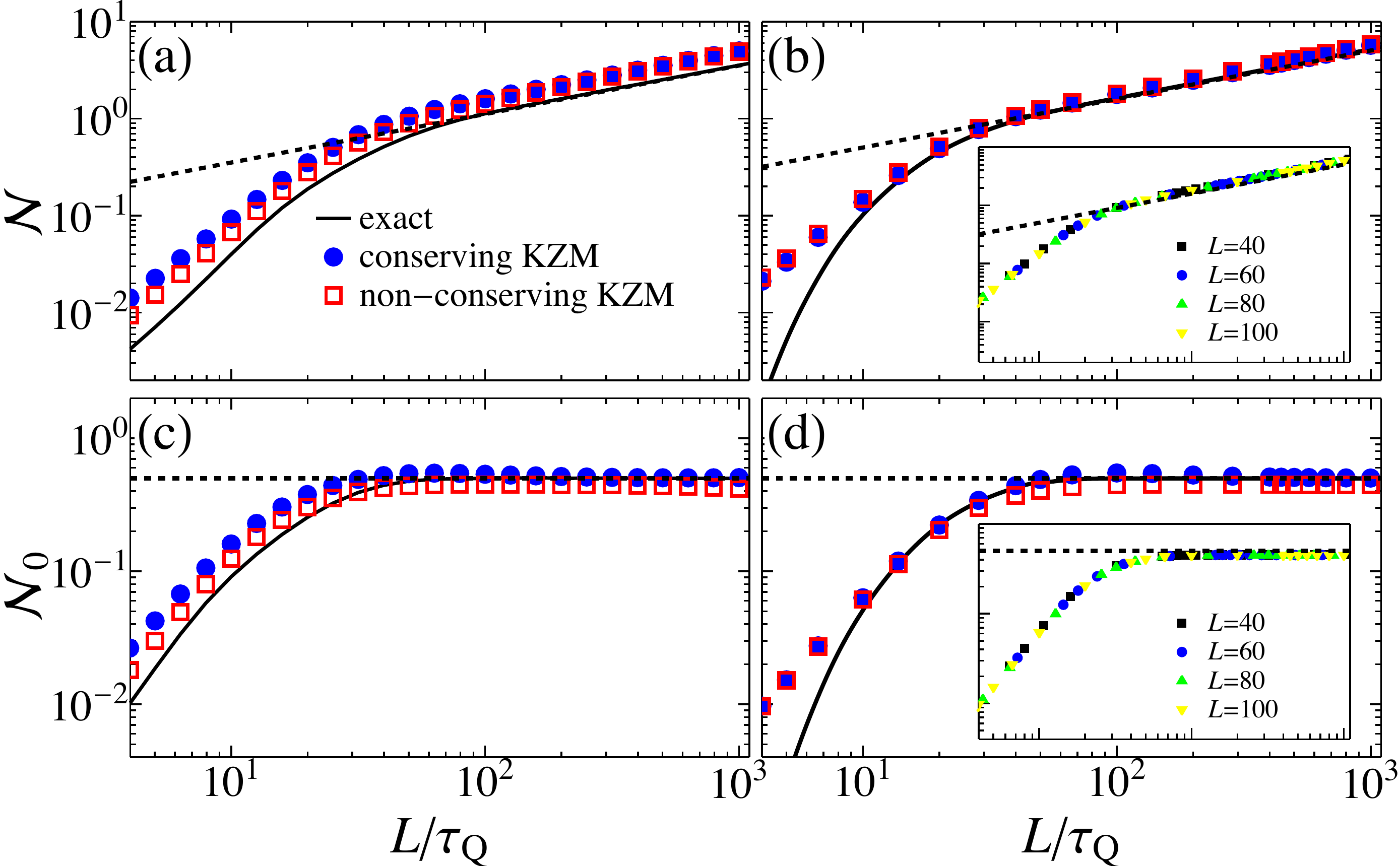}
\caption{(color online) The total number $\calN$ of defects (a,b) and the
  contribution $\calN_0$ of the Majorana state $\Phi_0$ (c,d) in the continuum
  model (a,c) and the lattice model (b,d). The black solid line, blue filled
  circle and red empty square are the results of the exact calculation,
  conserving and non-conserving KZM, respectively. The dashed line in (a) and
  (b) indicates the traditional KZM scaling $\calN\sim\sqrt{L/\tau_Q}$. In (a)
  and (c), $N=13$, and in (b) and (d), $L=100$. The insets illustrate the
  finite-size scaling for different sizes (calculated by the non-conserving
  KZM).}
\label{Paper:fig:2}
\end{figure}

\paragraph{Quasi-particle excitations.}

Let us now consider the dynamics of the \emph{many-body} Hamiltonian and apply the generalized KZM developed above.  Initially
($t_0=-\infty$) the system is prepared in its instantaneous many-body \emph{ground state} $\ket{G(t_0)}$, which is the vacuum of all \emph{positive-energy modes}, $\Phi_{2j+1}(x,t_0)$ and $\tau_x\Phi_{2j}^*(x,t_0)$:
\begin{equation}
\label{Paper::eq:14}
\hata_{2j+1}(t_0\ket{G(t_0)} =
\hatb_{2j}(t_0)\ket{G(t_0)} = 0;
\end{equation}
and in which all \emph{negative-energy modes}, $\Phi_{2j}(x,t_0)$ and
$\tau_x\Phi_{2j+1}^*(x,t_0)$, are occupied:
\begin{multline}
\label{Paper::eq:13}
\braket{G(t_0)|\hata_{2j}^\dag(t_0)\hata_{2j}|G(t_0)} \\{} =
\braket{G(t_0)|\hatb_{2j+1}^\dag\hatb_{2j+1}|G(t_0)} = 1 \,.
\end{multline}
We examine the number of excited quasi-particles $\calN$ in the far future ($t=\infty$). $\calN$ is directly related to the number of topological defects created by the quenching process across the critical point \cite{Dziarmaga05a}.
Due to the initial conditions~(\ref{Paper::eq:14}) and (\ref{Paper::eq:13}),
the occupancy of \emph{positive-energy modes},
$\tau_x\Phi_{2i}^*(x,\infty)$ and $\Phi_{2i+1}(x,\infty)$, are given by
$\sum_{j=0}^\infty P_{2i|2j+1}$ and
$\sum_{j=0}^\infty P_{2i+1|2j}$, respectively.
The total number $\calN$ of \emph{excited} quasi-particles is therefore given by
\begin{math}
\calN = \sum_{i,j=0}^\infty
\left[P_{2i|2j+1} + P_{2i+1|2j}\right].
\end{math}

The contribution of the Majorana mode,
\begin{math}
\calN_0\equiv\sum_{m\in\text{odd}} P_{m|0},
\end{math}
is of particular interest as it is known to defy the traditional KZM~\cite{Bermudez09a,Bermudez10a}.
It is stressed that the Majorana-mode contribution $\calN_0$ can be measured experimentally \cite{endnote:5}:
Consider two different quenching procedures; one starting from the ground state
$\ket{G(t_0)}$ and the other starting with the Majorana mode excited
$\hatb_0^\dag(t_0)\ket{G(t_0)}=\hata_0(t_0)\ket{G(t_0)}$. We find that
$\calN_0$ is related to the difference $\Delta\calN$ in $\calN$ for these two
processes by $\calN_0=(1-\Delta\calN)/2$ since
$\Delta\calN = \sum_{m\in\text{even}}P_{m|0} - \sum_{m\in\text{odd}}P_{m|0}$
and $\sum_{m\in\text{odd}}P_{m|0}+\sum_{m\in\text{even}}P_{m|0}=1$.

Figure~\ref{Paper:fig:2} shows $\calN$ and $\calN_0$ as a universal function of $L/\tau_Q$ for both the continuum and lattice models.
It demonstrates that the generalized (both conserving and non-conserving) KZM predictions agree well with the exact results. More importantly, it reveals three more prominent features of the generalized KZM distinguished clearly from the traditional KZM:
(i) The agreement remains good far beyond the traditional KZM scaling region.
The celebrated scaling behavior $\calN\sim\sqrt{L/\tau_Q}$ (i.e.,
$L\sqrt{\tau_0/\tau_Q}$ in the natural units and for the traditional
$L$-independent definition of $\tau_Q$ \cite{endnote:4}) is known
\cite{Zurek05a,Dziarmaga05a,Dziarmaga10a} to be valid only for relatively fast quenching
[Fig.~\ref{Paper:fig:2}(a) and (b)]. For slower quenching the exact dynamics
and the traditional KZM do not agree any longer. To the contrary, the
generalized KZM works remarkably well even for slow quenching
[Fig.~\ref{Paper:fig:2}(a) and (b)].
(ii) The Majorana-mode contribution $\calN_0$ defies completely the traditional KZM, as first pointed out in Refs.~\cite{Bermudez09a,Bermudez10a}, whereas it is well explained by the generalized KZM [Fig.~\ref{Paper:fig:2}(c) and (d)]. Since the Majorana mode plays a key role in the topological PT, understanding its dynamics is vital. Its inherent bound-state character and multi-level structure are efficiently captured by the generalized KZM.
(iii) The saturation of $\calN_0$ to $1/2$ for fast quenching [Fig.~\ref{Paper:fig:2}(c) and (d)] is intimately related to the multi-level structure and the localization of the Majorana state, and Eq.~(\ref{Paper::eq:16}) provides a simple explanation: Let $\tau_Q^*$ be the quenching time such that $M(t_1)L=-1$; $L/\tau_Q^*\approx 10$. For $\tau_Q\ll\tau_Q^*$, $M(t_1)L\ll-1$. It means that for such fast quenching the Majorana mode $\Phi_0(x,t_1)$ is well localized and its overlap with any bulk state $\Phi_m(x,t_m')$ is the same independent of $m$. Hence
\begin{math}
\calN_0
=\sum_{m\in\text{odd}}\left|\braket{\Phi_m(t_m')|\Phi_0(t_1)}\right|^2
\approx 1/2
\end{math}
since $\sum_{m\in\text{odd}}P_{m|0}\approx\sum_{m\in\text{even}}P_{m|0}$ in
this condition.
For slower quenching ($\tau_Q\gg\tau_Q^*$), on the other hand, $M(t_n)L>-1$ for
all $n$; namely, by the time the impulse region is reached, the state
$\ket{\Phi_0(t_1)}$ loses the Majorana character and the above argument does
not hold any longer.

We finally note that Ref.~\cite{Bermudez10a} studied (numerically) a different parameter regime of the same system~(\ref{Paper::eq:1}). They kept $\mu=0$ and varied $w$ from $-\Delta$ to $\Delta$. However, the dynamics is essentially the same.  With $\mu=0$, Eq.~(\ref{Paper::eq:1}) is decomposed into two decoupled Majorana chains that have opposite effective Dirac masses, $M(t)=w-|\Delta|$ and $-M(t)$, but are identical otherwise. Explicit calculation indeed reproduces their results.

In conclusion, we have developed a generalized KZM, which agrees well with the exact dynamics in a wide range of quenching rate. In particular, it successfully describes the contribution of the Majorana mode to the quenching-induced topological defects, which is essential in the dynamics of the topological PT.

\bibliographystyle{apsrev}
\bibliography{Paper}

\begin{thebibliography}{24}
\bibitem[{end({\natexlab{a}})}]{endnote:4}
\bibinfo{note}{Here we have adopted $L$-dependent $\tau_Q$ to be consistent
  with the relativity convention, where time $t$ and length $L$ has the same
  dimension. Replacing $\tau_Q\to\tau_Q/L$ recovers the traditional
  $L$-independent definition of $\tau_Q$ such as in
  \cite{Zurek05a,Dziarmaga05a,Bermudez09a,Bermudez10a}.}
\bibitem[{end({\natexlab{b}})}]{endnote:3}
\bibinfo{note}{For our purposes it suffices that $M(x,t)>0$ for $|x|>L/2$, but
  for simplicity we put it to be infinitely large. One physical realization was
  suggested in \cite{Fu08a}. Below we also discuss Kitaev's original lattice
  model, which is closer to realistic systems.}
\bibitem[{end({\natexlab{c}})}]{endnote:1}
\bibinfo{note}{It was found that $\alpha=\pi/2$ in the two-level case
  \citep{Damski06a}.}
\end{thebibliography}

\end{document}